# Dependences on *RE* of Superconducting Properties of Transition Metal co-doped (Ca,*RE*)FeAs$_2$ with *RE* = La-Gd

H. Yakita[a*], H. Ogino[a], A. Sala[a,b,c], T. Okada[a], A. Yamamoto[a], K. Kishio[a], A. Iyo[b], H. Eisaki[b], J. Shimoyama[a]

[a]*Department of Applied Chemistry, The University of Tokyo, 7-3-1 Hongo, Bunkyo, Tokyo,113 8656, Japan*
[b]*National Institute of Advanced Industrial Science and Technology (AIST), Tsukuba, Ibaraki, 05-8565, Japan*
[c]*University of Genova and CNR-SPIN, 16146 Genova, Italy*

Abstract

Dependence of superconducting properties of (Ca,*RE*)(Fe,*TM*)As$_2$ [(Ca,*RE*)112, *TM*: Co, Ni)] on *RE* elements (*RE* = La-Gd) was systematically investigated. Improvement of superconducting properties by Co or Ni co-doping was observed for all (Ca,*RE*)112, which is similar to Co-co-doped (Ca,La)112 or (Ca,Pr)112. $T_c$ of Co-co-doped samples decreased from 38 K for *RE* = La to 29 K for *RE* = Gd with decreasing ionic radii of $RE^{3+}$. However, Co-co-doped (Ca,Eu)112 showed exceptionally low $T_c$ = 21 K probably due to the co-existence of $Eu^{3+}$ and $Eu^{2+}$ suggested by longer interlayer distance $d_{\text{Fe-Fe}}$ of (Ca,Eu)112 than other (Ca,*RE*)112.

**1. Introduction**

After the discovery of superconductivity in LaFeAs(O,F)[1], exploration of iron-based superconductors has been attempted extensively and various type of new superconductors, such as 11[2], 111[3] and 122[4] systems, were discovered. Searches for new iron-based superconductors have been still continuing, resulted in a recent discovery of new series of iron-based superconductors (Ca,*RE*)FeAs$_2$((Ca,*RE*)112)[5-9]. (Ca,*RE*)112 has several unique properties, such as monoclinic structure with space group of $P2_1$ or $P2_1/m$, and co-existence of $As^{-1}$ and $As^{-3}$ in the crystal lattice[5,6]. Moreover, anisotropy of (Ca,La)112 was reported to be 2.0-4.2[10], which is smaller than that of 1111 type superconductors, while interlayer distance between Fe layer ($d_{\text{Fe-Fe}}$) is longer in (Ca,La)112. More than 15 % substitution of La for Ca site is needed to form (Ca,La)112 phase, and $T_c$ decreases with increasing La content[11]. It suggests that (Ca,La)112 is basically in the overdoped state. Sala *et al.* reported *RE* dependence on $T_c$ in (Ca,*RE*)112 with nominal *RE* composition of 15 % for Ca. A decrease in $T_c$ with a decrease in ionic radii of $RE^{3+}$ was observed except for (Ca,Ce)112 which did not show superconductivity. Kudo *et al.* reported the $T_c$ of 47 K in Sb co-doped (Ca,La)112[12]. Enhancement of $T_c$ in Sb co-doped (Ca,*RE*)112 is achieved by optimization of As-Fe-As bond angle. In addition, we have recently reported Mn or Co-co-doping effects in (Ca,La)112 and (Ca,Pr)112[13]. Analyzed substitution levels of Co in Co-co-doped (Ca,*RE*)112 were close to the nominal composition, though *RE* composition were higher than those of nominal compositions. Sharp superconducting transition in (Ca,La)112 and enhancement of $T_c$ in Ca,Pr)112 was induced by Co-co-doping, while Mn co-doping strongly suppressed superconductivity. These results indicates co-doping of transition metals (*TM*), such as Co, is promising way to improve superconducting properties of (Ca,*RE*)112.

In this paper, we report *TM*(Co, Ni) co-doping effect for (Ca,*RE*)112 (*RE* = La-Gd) with low *RE* doping level. *RE* element dependence on structural parameters as well as superconducting properties of *TM* co-doped (Ca,*RE*)112 are systematically investigated.

*Corresponding author at Room 502, Building 9, 2-11-16, Yayoi, Bunkyo, Tokyo, Japan, 113-8656. Tel.:+81-3-5841-7766
*E-mail address:* 8757570603@mail.ecc.u-tokyo.ac.jp.

## 2. Experiment

Samples with nominal compositions of $(Ca_{0.9}RE_{0.1})FeAs_2$ and $(Ca_{1-x}RE_x)Fe_{0.97}TM_{0.03}As_2$ ($RE$ = La-Gd, $TM$ = Co, Ni, $x$ = 0.075, 0.1) were synthesized starting from FeAs(3N), As(4N), Ca(2N), $RE$As($RE$ = La, Pr, Nd, Sm, 3N), $RE$ metals($RE$ = Ce, Eu, Gd, 3N), CoAs(3N), Ni(3N) powders. Starting powders were mixed and pelletized in an argon-filled glove box, filled in boron nitride(BN) crucible, and sintered at 1050-1200°C for 1 h under a high pressure of 2 GPa using a wedge-type cubic-anvil high-pressure apparatus (RIKEN CAP-07). Constituent phases and lattice constants were studied by powder XRD measurements using a RIGAKU Ultima-IV diffractometer, and the intensity data were collected in the $2\theta$ range of 5-80° in increments of 0.02° using Cu K$_\alpha$ radiation. Microstructure observation and compositional analysis were performed using a scanning electron microscopy (SEM; Hitachi High-Technologies TM3000) equipped with energy dispersive X-ray spectrometry (EDX; Oxford Instruments SwiftED 3000). The magnetic susceptibility was measured by a SQUID magnetometer (Quantum Design MPMS-XL5s).

## 3. Results and discussion

Figures 1(a)-(c) show XRD patterns of $TM$-free or $TM$ co-doped (Ca,$RE$)112 samples for $RE$ = Ce, Nd, Sm, Eu and Gd. (Ca,$RE$)112 phase formed as main phase in all samples, and impurity FeAs were always detected. BN detected in some samples is contaminated from high pressure media. The intensity of 002 peaks (~17°) of Ce, Nd, Sm doped samples are higher or almost same compared to that of main peaks (~37°), probably due to plate-like (Ca,$RE$)112 crystals. Figure 1(d) shows a typical surface image of Co-co-doped (Ca,Sm)112 bulk sample. The bulk is composed by plate-like crystals. Similar images were obtained for other $RE$ doped samples.

Figures 2(a)-(d) show the temperature dependences of the zero-field cooled (ZFC) and field cooled (FC) magnetization curves of $TM$-free, Co or Ni co-doped (Ca,$RE$)112 ($RE$ = La-Gd) samples. Ni co-doping for (Ca,La)112 decreased $T_c$ from 38 K to 36 K, while superconducting transition became sharp similar to that of Co-co-doped samples[13].

On the other hand, increases in $T_c$ by Co-co-doping were observed in other (Ca,$RE$)112 samples. $T_c$ of Co-co-doped (Ca,$RE$)112 was 31, 36, 34, 33, 21 and 29 K for $RE$ = Ce, Pr, Nd, Sm, Eu and Gd, respectively. $T_c$ exceeds 30 K for $RE$ = La-Sm in spite of direct substitution of Co for Fe site. Ni co-doping also increased $T_c$ of (Ca,$RE$)112 but their $T_c$ were slightly lower compared to Co-co-doped samples. These results suggest that Ni co-doped samples are in overdoped state.

As shown in Figs. 2(a) and (d), increases of $T_c$ by $TM$ doping were also confirmed for (Ca,Ce)112 and (Ca,Eu)112, while their $T_c$ values were relatively low compared to the other (Ca,$RE$)112. These low $T_c$ in Ce and Eu doped samples are consistent with that of $RE$ 15% doped samples reported by Sala et al.[7].

The low $T_c$ of (Ca,Ce)112 might be due to higher Ce concentration in the samples than that of other $RE$ doped samples. To confirm the idea, Co 3% doped $(Ca_{0.925}RE_{0.075})112$($RE$ = La, Ce) were synthesized under same conditions. Figures 3(a)-(c) show XRD patterns and ZFC and FC magnetization curves of the samples. Co-co-doped (Ca,$RE$)112 phase formed as main phase in spite of low nominal doping level of $RE$ in both samples. Decreasing nominal Ce composition increased $T_c$ from 31 K to 35 K. On the other hand, $T_c$ did not change by decreasing La composition in Co-co-doped (Ca,La)112. These results suggest that the low $T_c$ of $(Ca_{0.9}Ce_{0.1})112$ and $(Ca_{0.85}Ce_{0.15})112$[7] is due to relatively higher Ce substitution level for Ca site in the resulting samples. Compositional analysis by EDX revealed that actual Ce composition at Ca site was 14.8 % in Ce 7.5% doped sample while it was 19.3% and 26% in Ce 10% and Ce 15%[7] doped samples, respectively. The high Ce composition levels for Ca site is possibly due to ionic radii of $Ce^{3+}$(1.143 Å: C.N. = 8), which is closer to that of $Ca^{2+}$(1.120 Å) than that of $La^{3+}$(1.160 Å).

Figure 4(a) shows dependence of $T_c$ on ionic radii of $RE^{3+}$ of $TM$-free and $TM$ co-doped (Ca,$RE$)112, including the data of $RE$ 15% doped samples[7]. For $TM$-free compounds, a trend that $T_c$ decrease according to decrease of ionic radii of $RE^{3+}$ was found except for Ce and Eu doped samples showing exceptionally low $T_c$. Same tendency was observed for Co or Ni co-doped samples. On the other hand, $T_c$ of the Co-co-doped $(Ca_{0.925}Ce_{0.075})112$ was close to that of Co-co-doped $(Ca_{0.9}La_{0.1})112$ and $(Ca_{0.9}Pr_{0.1})112$. Figure 4 (b) shows relationship between interlayer distance between Fe planes $d_{Fe-Fe}$ values analyzed from XRD results and ionic radii of $RE^{3+}$ of $TM$-free and Co 3% co-doped (Ca,$RE$)112 samples. The values of $d_{Fe-Fe}$ decreased following decrease of ionic radii of $RE^{3+}$. However, those of Eu doped samples were relatively large considering ionic radii of $Eu^{3+}$(1.066 Å: C.N. = 8). Exceptionally large $d_{Fe-Fe}$ in (Ca,Eu)112 was also reported in *RE* 15% doped samples by Sala et al.[7]. Co-existence of $Eu^{2+}$ (1.25 Å: C.N. = 8) and $Eu^{3+}$ (1.066 Å) can be considered as the reason for relatively low $T_c$ and the enlarged $d_{Fe-Fe}$ of (Ca,Eu)112. Stürzer et al. have reported *RE* (*RE* = Y, La-Nd, Sm-Lu) doping effect for $Ca_{10}(Pt_3As_8)(Fe_2As_2)_5$ (10-3-8)[14], which have similar crystal structure to (Ca,*RE*)112. In these compounds, larger unit cell than other *RE* doped 10-3-8 and absence of superconductivity were observed in Eu doped 10-3-8. These are similar to that of (Ca,Eu)112, though superconductivity above 12 K was observed in (Ca,Eu)112. In addition, $d_{Fe-Fe}$ values are decreased by Co-co-doping with 0.01-0.02 Å compared to *TM*-free samples. This result is similar to the decrease in *c*-axis length by Co doping in LaFeAsO[15] and CaFeAsH[16], indicating that Co-co-doping also accompanies electron doping in (Ca,*RE*)112. Further careful studies are needed to clarify the reason of relatively high $T_c$ by direct *TM* doping to Fe site which increases electron carrier in already electron overdoped (Ca,*RE*)112[11].

## 4. Summary

In summary, Co or Ni co-doped (Ca,*RE*)112 have been synthesized for *RE* = La-Gd and their superconducting properties were investigated. Co or Ni co-doping improved superconducting properties of all (Ca,*RE*)112 samples. $T_c$s exceeding 30 K were observed in Co-co-doped (Ca,*RE*)112(*RE* = La-Sm) in spite of direct *TM* doping to Fe site. $T_c$ was found to decrease with a decrease in ionic radii of $RE^{3+}$ except for Ce and Eu doped samples. In the case of (Ca,Ce)112, $T_c$ increased by decreasing nominal Ce composition and $T_c$ of Co-co-doped $(Ca_{0.925}Ce_{0.075})$112 almost followed the $T_c$ trend. On the other hand, exceptionally low $T_c$ and long $d_{Fe-Fe}$ was found in (Ca,Eu)112, suggesting the co-existence of $Eu^{2+}$ and $Eu^{3+}$. Slight decrease in $d_{Fe-Fe}$ by Co-co-doping was also confirmed.

## Acknowledgements

This work was partly supported by the JSPS KAKENHI Grant Number 26390045, Izumi Science and Technology Foundation, and European Union-Japan project SUPER-IRON (grant agreement No. 283204).

Fig.1. XRD patterns of (a) $(Ca_{0.9}RE_{0.1})FeAs_2$, (b) $(Ca_{0.9}RE_{0.1})Fe_{0.97}Co_{0.03}As_2$ and (c) $(Ca_{0.9}RE_{0.1})Fe_{0.97}Ni_{0.03}As_2$ ($RE$ = Ce, Nd, Sm, Eu, Gd) sintered for 1 h under 2 GPa, and (d) secondary electron image of Co-co-doped (Ca,Sm)112 bulk sample

Fig.2. Temperature dependence of the ZFC and FC magnetization curves of $(Ca_{0.9}RE_{0.1})FeAs_2$, $(Ca_{0.9}RE_{0.1})Fe_{0.97}Co_{0.03}As_2$ and $(Ca_{0.9}RE_{0.1})Fe_{0.97}Ni_{0.03}As_2$ with (a) $RE$ = La, Ce (b) $RE$ = Pr, Nd, (c) $RE$ = Sm, Gd and (d) $RE$ = Eu

Fig.3. (a) XRD patterns and temperature dependence of the ZFC and FC magnetization curves of (b) $(Ca_{1-x}La_x)Fe_{0.97}Co_{0.03}As_2$ and (c) $(Ca_{1-x}Ce_x)Fe_{0.97}Co_{0.03}As_2$ ($x$ = 0.075, 0.1)

Fig.4. (a) $T_c$ and (b) $d_{Fe-Fe}$ for the high pressure synthesized samples as a function of ionic radii of the $RE^{3+}$ ions in a coordination number (C. N. ) of 8

Figure 1

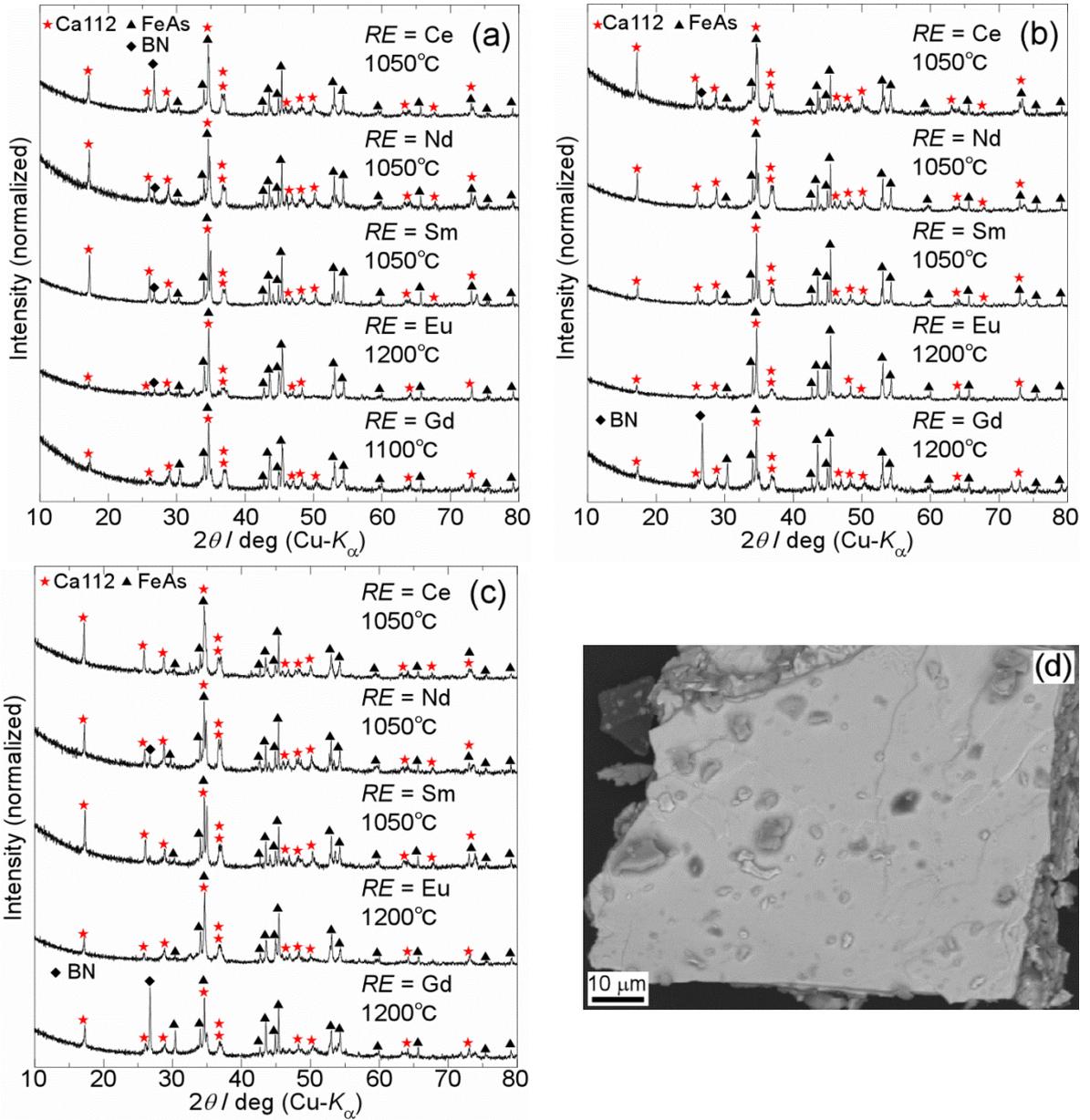

Figure 2

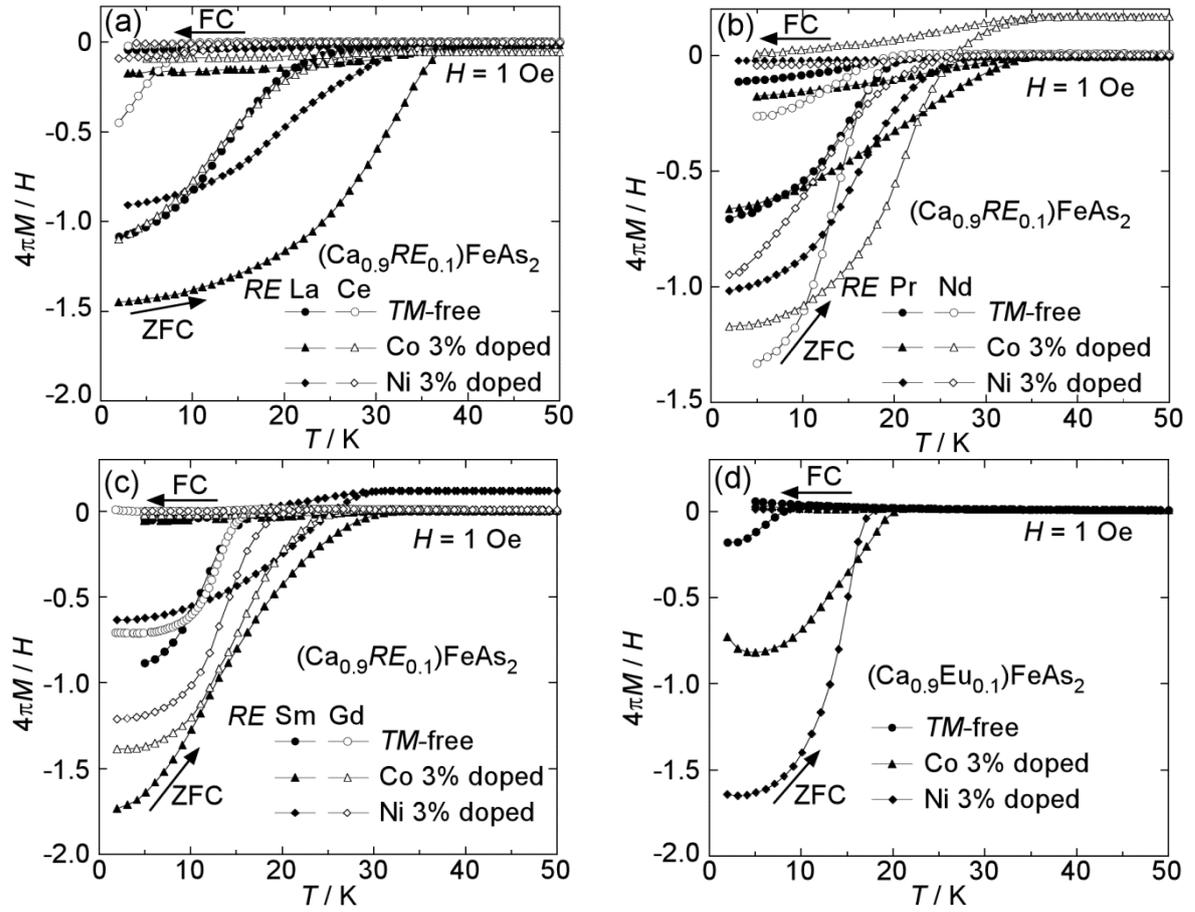

Figure 3

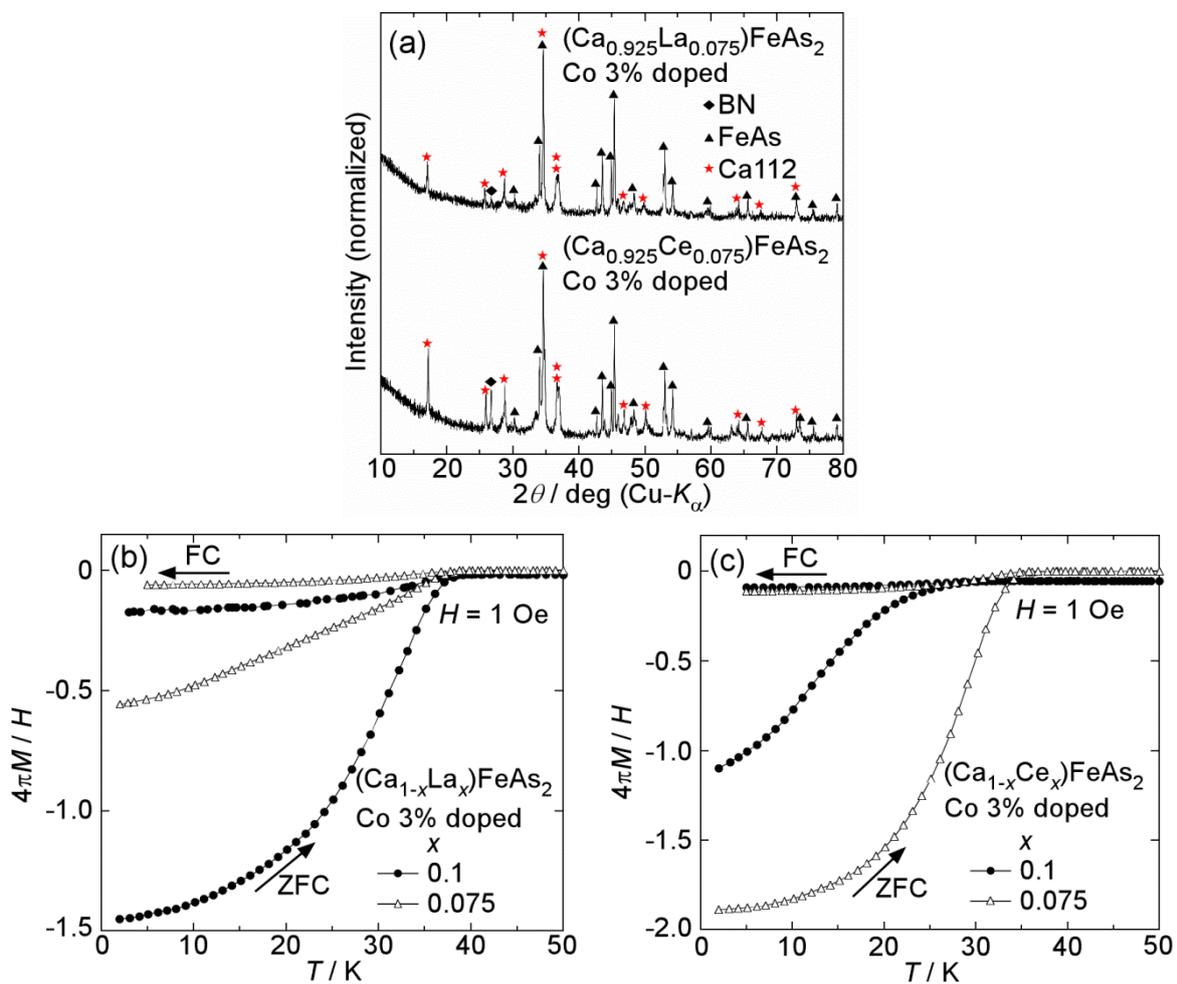

Figure4

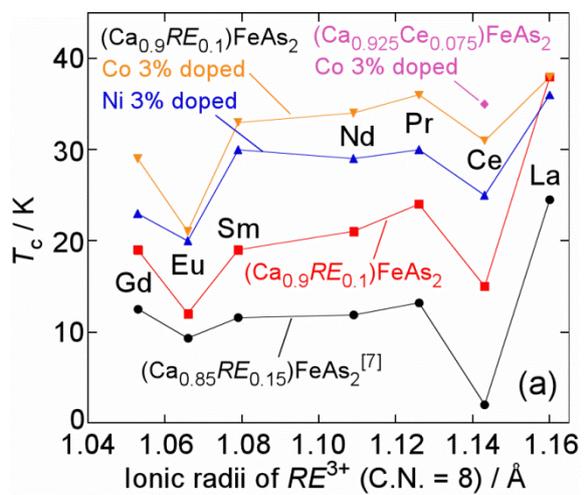 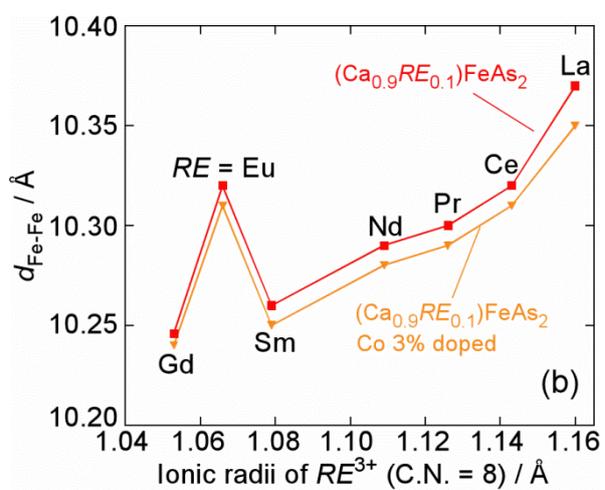